\documentclass[12pt]{elsart}
\usepackage{latexsym}
\usepackage{amsfonts,amssymb,amstext}
\renewcommand{\d}{\partial}

\newcommand{\C}{{\ensuremath{\mathbb C}}}

\newcommand{\M}{{\ensuremath{\mathbb M}}}

\newcommand{\x}{{\ensuremath{\mathbf x}}}
\newcommand{\z}{{\ensuremath{\mathbf z}}}
\newcommand{\bS}{{\ensuremath{\mathbf S}}}

\newcommand{\ee}{{\ensuremath{\mathbf e}}}
\newcommand{\f}{{\ensuremath{\mathbf f}}}
\newcommand{\p}{{\ensuremath{\mathbf p}}}
\newcommand{\X}{{\ensuremath{\mathsf X}}}
\newcommand{\E}{{\ensuremath{\mathsf E}}}
\newcommand{\g}{{\ensuremath{\mathbf g}}}

\newcommand{\y}{{\ensuremath{\mathbf y}}}

\newcommand{\dd}{{\ensuremath{\rm d}}}

\newcommand{\NPB}[3]{{\sl Nucl. Phys.} {\bf B#1} (#2) #3}

\newcommand{\MPLA}[3]{{\sl Mod. Phys. Lett.} {\bf A#1} (#2) #3}

\newcommand{\IJMP}[3]{{\sl Int. J. Mod. Phys.} {\bf A#1} (#2) #3}

\newcommand{\Tensor}[3]{{\sl Tensor, N.S.} {\bf #1} (#2) #3}
\begin{document}
\begin{frontmatter}
\title{Massive spinning particles and the geometry of null curves}
\author[Dubna]{Armen Nersessian\thanksref{emailar}}
and
\author[UAM]{Eduardo Ramos\thanksref{emailed}}

\address[Dubna]{Joint Institute for Nuclear Research\break
                Bogolyubov Laboratory of Theoretical Physics\break
                Dubna, Moscow Region, 141980, Russia}

\address[UAM]{Dept. de F\'{\i}sica Te\'orica, C-XI\break
                          Universidad Aut\'onoma de Madrid\break
                         Ciudad Universitaria de Cantoblanco\break
                          28049 Madrid, Spain}

\thanks[emailar]{\tt mailto:nerses@thsun1.jinr.ru}
\thanks[emailed]{\tt mailto:ramos@delta.ft.uam.es}

\begin{abstract}
We study the simplest geometrical particle model associated with null
paths in four-dimensional Minkowski space-time.  The action is given
by the pseudo-arclength of the
particle worldline. We show that the reduced classical phase space of
this system coincides with that of a massive spinning particle of spin
$s=\alpha^2/M$, where $M$ is the particle mass, and $\alpha$ is the
coupling constant in front of the action.  Consistency of the
associated quantum theory requires the spin $s$ to be an integer or
half integer number, thus implying a quantization condition on the
physical mass $M$ of the particle.  Then, standard quantization
techniques show that the corresponding Hilbert spaces are solution
spaces of the standard relativistic massive wave equations. Therefore
this geometrical particle model provides us with an unified
description of Dirac fermions ($s=1/2$) and massive higher spin
fields.
\end{abstract}
\end{frontmatter}

\section{Introduction}

The search for a classical particle model that under quantization
yields the Dirac equation and its higher spin generalizations has a
long history.  By far the most popular approach is to supersymmetrize
the standard relativistic particle model, whose action is given by the
proper time of the particle's path.  It is not difficult to show that
the system possessing $N=2s$ extended supersymmetry corresponds, after
quantization, to a massive spinning particle of spin
$s$. Nevertheless, the search for geometrical particle models of
purely bosonic character, which may lead to similar results, is
interesting on its own.  The reasons for this are quite simple, and
they may be exemplified by Polyakov's results on Fermi-Bose
transmutation in three dimensions. In \cite{PolyakovI}, Polyakov was
able to show how, in the context of three-dimensional Chern--Simons
theory, the presence of a torsion term in the effective action for a
Wilson loop was responsible for the appearance of Dirac fermions in an
otherwise apparently bosonic theory---thus opening the question of
which is the natural counterpart, if any, in the four-dimensional
case.

The purpose of this work is to show that there is a geometrical
particle model, based purely on the geometry of null curves in
four-dimensional Minkowski spacetime, that does the job, i.e., that
under quantization yields the wave equations corresponding to massive
spinning particles of arbitrary integer or half-integer
spin. The action for such a model is given by the pseudo-arclength,
the simplest among all the geometrical invariants associated with null
curves.

A few words are due now in order to motivate, from a more
`intuitive' point of view, our approach to this problem.

How is that a particle model based on null curves, which should
correspond a priori with massless particles, may be of relevance in
the massive case? The main clue that this can be indeed the case comes
from the {\em Zitterbewegung\/} associated with Dirac's equation.  It
is a well-known result that a measurement of the instantaneous
speed in the
relativistic theory of Dirac must lead to a result of $\pm c$---that
this is in no contradiction with Lorentz invariance is explained, for
example, in \cite{Dirac}---hence making plausible that light-like
paths may play a role in the corresponding particle system.

The second observation is a direct consequence of the one above: due
to the null character of the path any sensible local action should
depend on higher order derivatives. It is clear that a model for
spinning particles should contain extra degrees of freedom in order to
accommodate the ones associated with spin. Although some Lagrangian
models have been introduced in the literature where the extra degrees
of freedom are added by hand (see, e.g., \cite{spin} and references
therein), it would be highly desirable if they were provided by the
geometry itself. It is well known by now that higher derivative
theories may provide the required phase space. As we already
commented, that is the case in three dimensions, and moreover it has
also already proven to work in four dimensions, for massless
particles, with a Lagrangian given by the first curvature of the
particle's worldline \cite{rigid}.  Our choice has, then, been the
obvious one, this is to consider the simplest geometrical invariant
associated with null, or light-like, paths.  Therefore, and with no
further delay we will begin to introduce the necessary geometrical
background to work out the model at hand.

\section{Frenet equations for null curves in $\M^4$}

Let us start by constructing the Frenet equations associated with
null curves in four-dimensional Minkowski space-time. Our conventions
for the signature of the metric are $(+,-,-,-)$, i.e., time-like
vectors have a positive norm.

If we denote by $\x$ the embedding coordinate of the curve, the fact
that the curve is null implies that $\dot\x^2=0$. Let us denote by
$\ee_+=\dot\x$ the tangent vector to the curve. From $\ee_+^2=0$
it follows that there is a space-like vector $\ee_1$ such that
\begin{equation}
\dot\ee_+ =\sigma\ee_1,
\end{equation}
with $\ee_+\ee_1=0$, and $\ee_1^2=-1$. Although, a priori, a term
proportional to $\ee_+$ may appear in the right hand side of the
above equation it is always possible to reabsorb it
by a redefinition of $\ee_1$.

It will now show convenient to chose a parametrization for which
$\sigma=1$. This is equivalent to demanding that $\ddot\x\ddot\x=-1$,
and this can always be achieved by a change of parametrization unless
$\ddot\x\ddot\x=0$, this last case being trivial will be excluded from
our considerations.  This parametrization corresponds to choosing as our
time parameter the pseudo-arclength, which is defined as
\begin{equation}
\sigma (t) =\int_{t_0}^{t}
\left(-\ddot\x (t')\ddot\x (t') \right)^{1\over 4} dt'.
\end{equation}

We now pass to obtain Frenet equations in this parametrization. From
the definition of $\ee_1$ it follows that $\dot\ee_1\ee_1=0$ and
$\dot\ee_1\ee_+=1$, hence
\begin{equation}
\dot\ee_1 = A \ee_+ +\f_- + C\g_2,
\end{equation}
where the two new vectors $\g_2$ and $\f_-$ are chosen to obey that
$\g_2^2=-1$, $\f_-^2=0$, $\g_2\ee_1=\g_2\ee_+=\g_2\f_-= \ee_1\f_-=0$,
together with $\ee_+\f_-=1$.  From the fact that
$(\ee_+,\f_-,\ee_1,\g_2)$ span the tangent space at any point on the
curve the above equation trivially follows.

But notice that one can choose a
basis $(\ee_+,\ee_-,\ee_1,\ee_2)$ with
\begin{eqnarray}
\ee_- =&&\f_- + C\ee_2 +{1\over 2} C^2 \ee_+,\\
\ee_2 =&&\g_2 + C\ee_+
\end{eqnarray}
that, while preserving the same orthogonality relationships,
simplifies the above equation to yield
\begin{equation}
\dot\ee_1 = \kappa_1\ee_+ + \ee_-,
\end{equation}
with $\kappa_1= A-C^2/2$. The remaining Frenet equations
associated with the Frenet frame
$(\ee_+,\ee_-,\ee_1,\ee_2)$ follow
from the orthogonality relations, and the whole set reads
\begin{eqnarray}
\dot\x =&&\ee_+,\\
\dot\ee_+=&&\ee_1,\\
\dot\ee_1=&&\kappa_1\ee_+ + \ee_-,\\
\dot\ee_-=&&\kappa_1\ee_1 +\kappa_2\ee_2,\\
\dot\ee_2=&&\kappa_2\ee_+ .
\end{eqnarray}
Notice that in this case there are only two independent curvature
functions $\kappa_1$ and $\kappa_2$. This is intuitively obvious due
to the extra constraint on null curves coming from $\dot\x\dot\x=0$.

\section{The classical model}

We will consider the simplest geometrical action associated
with null curves, this is, its pseudo-arclength:
\begin{equation}
S =2\alpha\int \dd\sigma.
\end{equation}
It is convenient, although not strictly necessary, to
write the action in first order form. One then can write
\begin{equation}
S = \int dt
\left( 2\alpha\sqrt{-\dot\ee_+\dot\ee_+} +
\p (\dot\x -\ee_+\sqrt{-\dot\ee_+\dot\ee_+}) -
\lambda\ee_+^2\right).
\end{equation}
Notice that the equation of motion for the Lagrange
multipliers $\p$ and $\lambda$
imply that
\begin{equation}
\sqrt{-\dot\ee_+\dot\ee_+} = (-\ddot\x\ddot\x)^{1\over 4},
\end{equation}
thus proving the equivalence of both actions.

It is now straightforward to develop the Hamiltonian formalism
following Dirac's description for singular Lagrangians. We will skip
the details, which are completely standard.  After a little work one
arrives to a symplectic form
\begin{equation}
\Omega = \dd\p\wedge\dd\x +\dd\p_+\wedge\dd\ee_+,
\end{equation}
endowed with the following set of primary constraints
\begin{eqnarray}
\phi_1=&&\p_+^2 + (\p\ee_+ -2\alpha)^2,\\
\phi_2=&& \ee_+^2,
\end{eqnarray}
and secondary ones
\begin{eqnarray}
\phi_3=&&\p_+\ee_+,\\
\phi_4=&&\p\ee_+ -\alpha,\\
\phi_5=&&\p\p_+.
\end{eqnarray}
The Hamiltonian is of the form
\begin{equation}
H= v ( \p_+^2 +\p^2\ee_+^2 + (\p\ee_+ -2\alpha)^2 ),
\end{equation}
where $v$ should be regarded as an arbitrary function that is
only fixed by choosing a particular parametrization.
If one chooses to parametrize the path by pseudo-arclength
one gets that $v=-1/2\alpha$.

There is only one first class constraint,
the generator of reparametrizations. The dimension of the reduced phase
space is therefore $10$.

The equations of motion, in the pseudo-arclength parametrization,
are given by
\begin{eqnarray}
\dot\x =&&\ee_+,\\
\dot\ee_+=&&-{1\over\alpha}\p_+,\\
\dot\p_+=&&{1\over\alpha}\p^2\ee_+ -\p,\\
\dot\p=&&0.
\end{eqnarray}

Consistency of these equations of motion with Frenet equations
imply that $\p_+=-\alpha\ee_1$, $\p =\alpha\ee_- +
{\p^2}\ee_+/2\alpha$,
together with
\begin{equation}
\kappa_1 =-{1\over 2\alpha^2}\p^2.
\end{equation}
The solution of these equations of motion are particular
examples of null helices \cite{Bonnor}.

The key observation, in order to arrive to
a manageable expressions for
the symplectic form on the
constrained surface, is that it is possible to define a
standard free coordinate $\X$ out from the canonical variables
of our model.
Notice that
\begin{equation}
\X =\x -{\alpha\over {\p^2}}\p_+
\end{equation}
has the property that its time derivative is given by
\begin{equation}
\dot\X =\dot\x -{\alpha\over {\p^2}}\dot\p_+ =
{\alpha\over {\p^2}}\p.
\end{equation}
And from this it trivially follows that $\ddot\X=0$.

It is then natural to introduce the new coordinate
\begin{equation}
\E_+ =\ee_+ -{\alpha\over {\p^2}}\p
\end{equation}
so that the symplectic form $\Omega$ takes the simple form
\begin{equation}
\Omega=\dd\p\wedge\dd\X + \dd\p_+\wedge\dd\E_+
\end{equation}
on the constrained surface.

The constraints may be equally expressed in these variables, and they
read
\begin{eqnarray}
\p_+^2 +\alpha^2&&= 0,\\
\p_+\E_+&&= 0,\\
\E_+^2 + {\alpha^2\over {\p^2}}&&= 0,\\
\p\p_+&&= 0,\\
\p\E_+&&= 0.
\end{eqnarray}

This constraint system suggests the introduction of the following
complex coordinates
\begin{equation}
\z=\p_+ +i \sqrt{\p^2}\,\E_+.
\end{equation}
In terms of $\z$ the constraints simply read
\begin{equation}
\z^2= 0, \quad \z\bar\z + 2\alpha^2= 0,\quad
{\rm and}\quad \p\z= 0.
\end{equation}

Let us recall that the irreducible representations of the
Poincar\'e algebra are labeled by the values of the two Casimirs
$\p^2$ and the square of the Pauli-Lubansky vector
\begin{equation}
S^{\mu} ={1\over 2}\epsilon^{\mu\nu\rho\sigma}p_{\nu}M_{\rho\sigma},
\end{equation}
with $M_{\rho\sigma}$ the generator of Lorentz transformations.
In our particular case the explicit expression for $S_{\mu}$ reads
\begin{equation}
S_{\mu} =\epsilon_{\mu\nu\rho\sigma}p^{\nu}p_+^{\rho}\e_+^{\sigma},
\end{equation}
and one obtains that $\bS^2=-\alpha^4$, while there is no
restriction on the possible values of $\p^2$. This implies that our
phase space is not elementary, i.e., the Poincar\'e group does
not act in a transitive way. In order to obtain irreducible
representations of the Poincar\'e group under quantization--
physical states should always be decomposable into irreducible
representations of the Poincar\'e group-- we will study the elementary
phase spaces defined through the extra constraint
\begin{equation}
\p^2 = M^2,
\end{equation}
with $M$ a free parameter with dimensions of mass. A priori $M^2$
could be positive, negative or zero, we will only consider the first
case, because the other two correspond to unphysical irreducible
representations of the Poincar\'e group, i.e., tachyonic and continuous
spin representations, respectively.

Because of the null character of $\z$ it will show convenient
to introduce a spinor parametrization of the reduced phase space.
Given an arbitrary complex four-vector $\y$ it can be rewritten in
spinor coordinates as follows:
\begin{equation}
(y^{A\dot A})=
{1\over \sqrt{2}}
\left(
\begin{array}{cc}
y^0 +y^3&y^1+iy^2\\
y^1-iy^2&y^0-y^3
\end{array}
\right)
\end{equation}
so that ${\rm det}(y^{A\dot A})={1\over 2} g_{\mu\nu}
y^{\mu}y^{\nu}$.

Because of the two-to-one local isomorphism between
$SL(2,\C)$ and the identity component of the Lorentz
group, one such Lorentz transformation on $\y$ is
equivalently represented by the action of an $SL(2,\C)$
matrix acting on the undotted indices and its
complex conjugate matrix on the dotted ones. Raising
and lowering of indices is mimicked in spinor language
by contraction with the invariant antisymmetric tensors
$(\epsilon^{AB})=(\epsilon_{AB})$, with $\epsilon^{01}=+1$,
and analogous expressions for the dotted indices.

The null character of $\z$ now implies that
\begin{equation}
z^{A\dot A} =\sqrt{2}\alpha \xi^A\bar\eta^{\dot A},
\end{equation}
and from the second of our constraints it must follow that
$\xi^A\eta_A=1$, or equivalently that $\xi$ and $\eta$ form
an spinor basis.
Notice, though, that this does not completely fix the spinors $\xi$
and $\eta$ because one still has the freedom to rescale both as
follows
\begin{equation}
\xi^A\rightarrow a \,\xi^A\quad{\rm and}\quad
\eta^A\rightarrow {1\over a}\eta^A,
\end{equation}
with $a$ an arbitrary (nonzero) real number. This residual freedom
will be fixed as follows. Let us consider the remaining constraint
$\p\z= 0$. In spinor coordinates it reads
\begin{equation}
p^{A\dot A}\xi_A\bar\eta_{\dot A}= 0
\end{equation}
or equivalently
\begin{equation}
p^{A\dot A}\xi_A =\Lambda\sqrt{\p^2}\bar\eta^{\dot A}=
\Lambda M \bar\eta^{\dot A},
\end{equation}
with $\Lambda$ an arbitrary (nonzero) real number. One then may
fix completely the freedom to rescale the spinors by setting
$\Lambda =\pm 1/\sqrt{2}$, and then one may finally write the
only remaining constraint in the following form
\begin{equation}
\p^{A\dot A}\xi_A\bar\xi_{\dot A} =\pm{M\over \sqrt{2}}.\label{Klein}
\end{equation}
A direct computation now shows that the symplectic form
in spinor coordinates reads
\begin{equation}
\Omega =\dd\p\wedge\dd\X \pm i {\sqrt{2} s\over M}
(p_{A\dot A}\dd\xi^A\wedge\dd\bar\xi^{\dot A} +
\bar\xi^{\dot A} \dd\xi^A\wedge \dd p_{A\dot A})
+\quad c.c.,
\end{equation}
where $c.c.$ stands for complex conjugate of the previous term,
and $s$ is a dimensionless parameter defined as
\begin{equation}
s={\alpha^2\over M}.
\end{equation}

\section{The quantum theory}

One may now check (see \cite{Woodhouse}) that the
reduced phase space can be identified with the coadjoint
orbit of the Poincar\'e group associated with a representation of
mass $M$ and spin $s$. Quantization is then a standard exercise whose
solution can be found, for example, in the excellent book on geometric
quantization by Woodhouse \cite{Woodhouse}.
Nevertheless, because of completeness,
and the desire to simplify things a little to the less mathematically
oriented reader, we will now sketch how the quantization procedure
may be carried away in spinor coordinates. In particular, we will show
how Dirac equation may be obtained in the $s=1/2$ case.

One should start by choosing a polarization, or in simpler words
one
should demand the wave function to depend
only on half of the canonical coordinates. Our choice will be that
the wave function will depend on $\p$ and $\xi$.

In order to quantize the system we should implement the remaining
first class constraints
as conditions on the wave function.
The first constraint $\p^2=M^2$ implies that the wave function
has support on the mass hyperboloid. In order to implement
the second one (\ref{Klein})
notice that, roughly speaking,
$p_{A\dot A}\bar\xi^{\dot A}$ and $\xi$ are conjugate variables
in the induced symplectic form, therefore under quantization
\begin{equation}
p_{A\dot A}\bar\xi^{\dot A}\rightarrow{\d\ \over\d\xi^A}
\end{equation}
up to some numerical factors. A careful computation shows that
(\ref{Klein}) implies at the quantum level that
\begin{equation}
\xi^A{\d\psi(\p,\xi )\over\d\xi^A}= 2s \psi(\p,\xi),
\end{equation}
i.e., the wave function is a homogeneous function of $\xi$ of degree
$2s$. Then single-valuedness of our wave function under
$\xi\rightarrow {\rm exp}(2\pi i)\xi$ requires $s$ to be an integer or
half-integer number. Therefore one has that
\begin{equation}
\psi (\p,\xi )=\psi_{A_1 A_2\cdots A_{2s}}(\p)\xi^{A_1}\xi^{A_2}
\cdots\xi^{A_{2s}},
\end{equation}
where $\psi_{A_1 A_2\cdots A_{2s}}(\p)$ has support on the mass
hyperboloid.
It is now straightforward to check that
its Fourier transform
\begin{equation}
\varphi_{A_1 A_2\cdots A_{2s}}(\x) =
\left( {1\over 2\pi}\right)^{3\over 2}
\int_{H_M^s}\psi(\p)_{A_1 A_2\cdots A_{2s}} {\rm e}^{-i\p\x}
{\rm d}\tau,
\end{equation}
with $H_M^s$ the positive energy branch of the mass hyperboloid, and
${\rm d}\tau$ its invariant measure, yield positive frequency
solutions of the massive wave equation
\begin{equation}
\Box\varphi_{A_1 A_2\cdots A_{2s}}(\x) +
M^2\varphi_{A_1 A_2\cdots A_{2s}}(\x)=0.\label{Gordon}
\end{equation}
Notice that, for example, for $s=1/2$ this equation has the same physical
content that the Dirac equation. This is so because the dotted component
of the four component Dirac spinor, which will be denoted by
$\chi_{\dot A}$, may
be defined trough the relation
\begin{equation}
\nabla_{A\dot A}\varphi^{A} = {M\over \sqrt{2}}\chi_{\dot A}.
\end{equation}
This together with (\ref{Gordon}) implies that
\begin{equation}
\nabla_{A\dot A}\chi^{\dot A} = {M\over \sqrt{2}}\varphi_{A},
\end{equation}
thus being equivalent to the standard Dirac equation for $(\varphi,
\chi)$.

Although this short discussion about the quantization of the system
cannot make justice to this extensive topic, the interested reader 
may fill the gaps with the help of the current literature on the
subject.

\begin{ack}
E. Ramos would like to thank J. Roca and J.M. Figueroa-O'Farrill
for many useful conversations on
the subject.  The work of A.N. has been partially supported by grants
INTAS-RFBR No.95-0829, INTAS-96-538 and INTAS-93-127-ext.

\end{ack}

\end{document}